# The effect of anisotropy on the absorption spectrum and the density of states of two-dimensional Frenkel exciton systems with Gaussian diagonal disorder

I. Avgin[a] and D. L. Huber[b,*]

[a] Department of Electrical and Electronic Engineering, Ege University, Bornova 3500, Izmir, Turkey

[b] Physics Department, University of Wisconsin-Madison, Madison, WI 53706, USA

**Abstract**

Using the coherent potential approximation, we investigate the effects of anisotropy and disorder on the optical absorption and the density of states of Frenkel exciton systems on square, rectangular, and triangular lattices with nearest-neighbor interactions and a Gaussian distribution of transition frequencies. The analysis is based on an elliptic integral approach that gives results over the entire spectrum. It is found that the absorption is weakly affected by the anisotropy in contrast to the density of states where the effects can be much stronger. The results for the square lattice are in good agreement with the finite array calculations of Schreiber and Toyozawa. Our findings suggest that the coherent potential approximation can be useful in interpreting the optical properties of two-dimensional systems with dominant nearest-neighbor interactions and Gaussian diagonal disorder where the optically excited states are Frenkel excitons.

Keywords: Frenkel exciton; square, rectangular and triangular lattices; diagonal disorder, Gaussian fluctuations; coherent potential approximation

[*] Correspondence to Physics Dept., Univ. of Wisconsin-Madison, 1150 University Ave., Madison, WI 53706, USA

E-mail address: dhuber@wisc.edu (D. L. Huber)

# 1. Introduction

In a series of recent papers, [1-4], we have applied the coherent potential approximation (CPA) to the calculation of the optical absorption and density of states of the Frenkel exciton model in one, two and three dimensional arrays with nearest-neighbor interactions and Gaussian disorder associated with the single-site transition frequencies. In the case of the one-dimensional systems, we have shown that the results for the density of states are in excellent agreement with numerical calculations carried out on arrays of $10^7$-$10^8$ sites [1]. The accuracy of the CPA for the optical absorption in one dimension [2] was tested in a comparison with data obtained from finite-array calculations by Schreiber [5]. Good agreement was obtained with the ensemble average of data from arrays of 199 sites. The CPA was applied to cubic lattices in [3]. In the case of the simple cubic lattice, reasonable agreement was obtained with the corresponding finite array calculations of Toyozawa and Schreiber [6,7] for the optical absorption and the density of states. Along with the simple cubic data, corresponding CPA results were reported for both the body-centered and face-centered cubic lattices.

A preliminary CPA analysis of the optical absorption and the density of states for the square lattice was reported in [4]. The approach followed was based on a large-energy expansion of the Green's function, and the calculations were limited to energies below the absorption edge of the ideal system. Like the simple cubic analysis, the results were in reasonable agreement with findings reported in [6,7]. In this paper, we investigate the square lattice using an approach for the calculation of the Green's function that is based on the numerical evaluation of a complete elliptic integral of the first kind with complex modulus. Unlike the previous approach, we obtain results that are applicable over the entire absorption band. We also extend the theory to rectangular and triangular lattices for which there are as yet no finite-array results to compare with. The main focus of this paper is on the effects of interaction anisotropy on the absorption and the density of states in disordered Frenkel systems.

The density of states, $\rho(E)$, is proportional to the number of exciton modes with energies between $E$ and $E + dE$. We take it to be normalized, i.e.

$$\int_{-\infty}^{\infty} dE \rho(E) = 1. \tag{1}$$

We also introduce a dimensionless optical absorption, $F(E)$, or lineshape function, defined by the sum [2,6]

$$F(E) = N^{-1} \sum_{\mu} |\sum_{n=1}^{N} \psi_n^{\mu}|^2 \delta(E - \varepsilon_{\mu}) \tag{2}$$

where $\varepsilon_{\mu}$ denotes the energy of exciton mode $\mu$, and $\psi_n^{\mu}$ is the normalized amplitude of mode $\mu$ on lattice site $n$. Similar to the density of states, we have

$$\int_{-\infty}^{\infty} F(E) dE = 1 \tag{3}$$

In the CPA, the starting point in all three dimensions is the lattice Green's function which in turn depends on the structure of the lattice and the exciton energy. The calculation of the optical absorption and the density of states follows once one knows the CPA Green's function which is obtained from the ideal lattice Green's function, $G_0(E)$, by the replacement $E \rightarrow E - V_C(E)$, where $V_C(E)$ is the coherent potential. For the CPA density of states, we have

$$\rho(E) = \pi^{-1} \operatorname{Im} G_0(E - V_C(E)) \tag{4}$$

in which Im denotes the imaginary part, while for the CPA absorption, one has [2]

$$F(E) = \pi^{-1} \operatorname{Im}([E - V_C(E) - E_{\mathbf{k}=0}]^{-1}), \tag{5}$$

where $E_{\mathbf{k}=0}$ is the exciton energy at the band edge. Equations relating to the calculation of $V_C(E)$ are given in [1-4] and will not be reproduced here.

It should be mentioned that in addition to optically active monolayers, there are quasi-two dimensional magnetic exciton systems where the nearest-neighbor approximation is appropriate. Among these are $Rb_2MnF_4$ and $K_2NiF_4$ where there is a very weak coupling between the magnetic ions in different planes, and the in-plane Mn and Ni sites form a square lattice.

## 2. Square lattice

In analyzing the square and rectangular systems, we take the unit of energy to be the width of the ideal (no disorder) exciton band and assume the absorption edge is at the bottom of the band. In the case of the square lattice, this leads to the exciton energy

$$E_{\mathbf{k}} = -(1/4)(\cos k_x + \cos k_y) \tag{6}$$

in the absence of disorder. The corresponding expression for the ideal lattice Green's function for complex energy takes the form [8].

$$G_0(E) = (2/\pi E) \mathbf{K}(2E^{-1}) \tag{7}$$

where $\mathbf{K}(m)$ denotes the complete elliptic integral with modulus $m$.

The Gaussian averaging associated with the diagonal disorder is discussed in detail in [2-4]. It depends on the variance, $\sigma^2$. Following [4-6], we carried out calculations for the square and triangular lattices with $\sigma^2 = 0.263, 0.186, 0.131, 0.093$, and $0.066$ where $\sigma$ is in units of the ideal band width. Our results for the normalized optical absorption, $F(E)$, and the normalized density of states, $\rho(E)$, for the square lattice are shown in Figs. 1 and 2, respectively. They are seen to be in good agreement with the finite-array calculations of Schreiber and Toyozawa as displayed in Figs. 2 [5] (absorption) and 5(b) [6] (density of states). This result is also consistent with [4], where a similar level of agreement was established for $E < -0.5$.

## 3. Rectangular lattice

In the absence of disorder, the exciton energy for a rectangular lattice with unit band width takes the form

$$E_{\mathbf{k}} = -(\lambda \cos k_x + \cos k_y)/2(\lambda+1), \quad \lambda > 0 \tag{8}$$

The corresponding expression for the lattice Green's function is given in the Appendix to [8]

$$G_0(E) = 2(\lambda+1)(\pi\lambda^{1/2})^{-1} k_1(E) \mathbf{K}(k_1(E)) \tag{9}$$

where

$$k_1(E) = \{4\lambda/[4(\lambda+1)^2 E^2 - (\lambda-1)^2]\}^{1/2} \tag{10}$$

Since the absorption is weakly affected by changes in $\lambda$, we show the only results for the density of states where we take $\sigma^2 = 0.131$ and $\lambda = 0.5, 1.0$ and $1.5$ (Fig. 3). It is apparent that the density of states near the center of the band is strongly affected by changes in $\lambda$. The lower curve shows the results for $\lambda = 0.5$, the middle curve for $\lambda = 1.5$ and the upper curve for $\lambda = 1.0$. The dip at the center of the band, which grows stronger as with increasing anisotropy, reflects the fact that in the limits $\lambda \gg 1$ and $\lambda \to 0$, the system approaches an array of weakly coupled chains with the consequence that the optical properties and the density of states become characteristic of one-dimensional arrays where there are singularities at $E = \pm 1/2$ [1,2].

## 4. Triangular lattices

In the absence of disorder, the exciton energy for the isotropic triangular lattice with unit interaction strength has the form [9]

$$E_{\mathbf{k}} = -(\cos 2k_x + 2\cos k_x \cos k_y) \tag{11}$$

with the optical absorption edge at $-3$ and the upper edge at $3/2$. Similar to the face-centered cubic lattice, and unlike the square and rectangular lattices, the exciton band does not have a reflection point on the real axis. In units of the band width, $9/2$, the corresponding lattice Green's function for complex energy with negative real part has the form

$$G_0(E) = (9/4\pi) g(E) \mathbf{K}(k_2(E)) \tag{12}$$

where

$$g(E) = -8/[(-9E+3)^{1/2} - 1]^{3/2} [(-9E+3)^{1/2} + 3]^{1/2} \tag{13}$$

and

$$k_2(E) = 4(-9E+3)^{1/4}/[(-9E+3)^{1/2} - 1]^{3/2} [(-9E+3)^{1/2} + 3]^{1/2} \tag{14}$$

In the triangular lattice discussed above all nearest-neighbor interactions are the same. In [10], Horiguchi investigated an anisotropic triangular lattice in which interactions along two of the connecting lines took on the value $\gamma$ while the interaction on the third line was equal to 1. When the optical edge is at the bottom of the band, the exciton energy is given by

$$E_{\mathbf{k}} = -[\cos(2ak_x) + 2\gamma \cos(ak_x)\cos(bk_y)] \tag{15}$$

where $a = 1/2$ and $b = \sqrt{3}/2$ when the length of the side of the triangle is 1 [10]. The lower and upper band edges and the band width (BW) are given in Table 1.

| Range of $\gamma$ | Lower band edge | Upper band edge | Band width (BW) |
|---|---|---|---|
| $0 < \gamma < 2$ | $-(1+2\gamma)$ | $1 + \gamma^2/2$ | $2 + 2\gamma + \gamma^2/2$ |
| $\gamma = 2$ | $-5$ | $3$ | $8$ |
| $\gamma > 2$ | $-(1+2\gamma)$ | $2\gamma - 1$ | $4\gamma$ |

**Table 1**. The locations of the band edges and the band width for the anisotropic triangular lattice. The nearest-neighbor interaction on two of the lines of sites is equal to $\gamma$ and equal to 1 on the third line.

With the energy in units of the band width, the density of states is strongly affected by the ratio of the interaction strengths. This is evident in Fig. 4 which displays $\rho(E)$ in the absence of disorder for $\gamma = 1/2$, 1, and 3. The locations of the peaks can be understood from the analysis in [10]. The singularities in the ideal Green's function for the anisotropic triangular lattice are associated with $\gamma$-dependent critical points on the energy axis and corresponding peaks in the density of states. In units of the band width, BW, the peaks are at $(2\gamma - 1)$/BW and 1/BW for $0 < \gamma < 1$. For $\gamma = 1$, there is a single peak at 1/BW = 0.25. For $1 < \gamma < 2$, the peaks are also at $(2\gamma - 1)$/BW and 1/BW, and when $\gamma \geq 2$ there is a single peak at 1/BW (Fig. 4). In Figs. 5 – 10, we show the effects of disorder on the optical absorption and the density of states for $\gamma = 1/2$, 1, and 3. The behavior of the density of states for $\gamma \ll 1$ and $\gamma \gg 1$ is consistent with the fact that in the former limit, the triangular array is equivalent to an ensemble of non-interacting chains so the density of states approaches that of the one dimensional array [1,2], whereas in the latter limit, each site has four nearest neighbors with the same interactions and the density of states resembles that of the square lattice.

## IV. Discussion

The principal conclusion of this paper is that the introduction of anisotropy (square → rectangular; isotropic triangular → anisotropic triangular) has a pronounced effect on the distribution of modes in two-dimensional disordered Frenkel exciton systems with nearest-neighbor interactions, whereas it has a much weaker effect on the optical absorption. It was found that in the rectangular and triangular lattices the qualitative features of the absorption are only weakly affected by the anisotropy for a given value of $\sigma$. In contrast, the density of states is strongly affected. This difference reflects the fact that the absorption is related to the response of the system to a uniform ($k = 0$) electromagnetic field whereas the density of states contains contributions from all ***k*** values [2].

**Figure captions**

**Fig. 1.** Normalized optical absorption for the square lattice. From left to right along the energy axis, the curves correspond to $\sigma^2 =$ 0.263, 0.186, 0.131, 0.093, and 0.066. In this and all other figures, energy is in units of the band width.

**Fig. 2.** Normalized density of states for the square lattice. From left to right along the energy axis, the curves correspond to $\sigma^2 =$ 0.263, 0.186, 0.131, 0.093, and 0.066.

**Fig. 3.** Normalized density of states for the rectangular lattice with $\sigma^2 =$ 0.131. From bottom to top at $E = 0$, the curves are for $\lambda =$ 0.5, 1.5 and 1.0.

**Fig. 4.** Normalized density of states for ideal (no disorder) triangular lattices with $\gamma = 1/2$ (red squares), 1(gray solid line), and 3 (black rectangles).

**Fig. 5.** Normalized optical absorption for the anisotropic ($\gamma = 1/2$) triangular lattice. From left to right along the energy axis, the curves correspond to $\sigma^2 =$ 0.263, 0.186, 0.131, 0.093, and 0.066.

**Fig. 6.** Normalized density of states for the anisotropic isotropic ($\gamma = 1/2$) triangular lattice. From left to right along the energy axis, the curves correspond to $\sigma^2 =$ 0.263, 0.186, 0.131, 0.093, and 0.066. Note that the density of states is not symmetric about $E = 0$.

**Fig. 7.** Normalized optical absorption for the isotropic triangular lattice with $\gamma = 1$. From left to right along the energy axis, the curves correspond to $\sigma^2 =$ 0.263, 0.186, 0.131, 0.093, and 0.066.

**Fig. 8.** Normalized density of states for the isotropic triangular lattice with $\gamma = 1$. From left to right along the energy axis, the curves correspond to $\sigma^2 =$ 0.263, 0.186, 0.131, 0.093, and 0.066.

**Fig. 9.** Normalized optical absorption for the anisotropic triangular lattice with $\gamma = 3$. From left to right along the energy axis, the curves correspond to $\sigma^2 =$ 0.263, 0.186, 0.131, 0.093, and 0.066. Energy is in units of the band width.

**Fig. 10.** Normalized density of states for the anisotropic triangular lattice with $\gamma = 3$. From left to right along the energy axis, the curves correspond to $\sigma^2 =$ 0.263, 0.186, 0.131, 0.093, and 0.066.

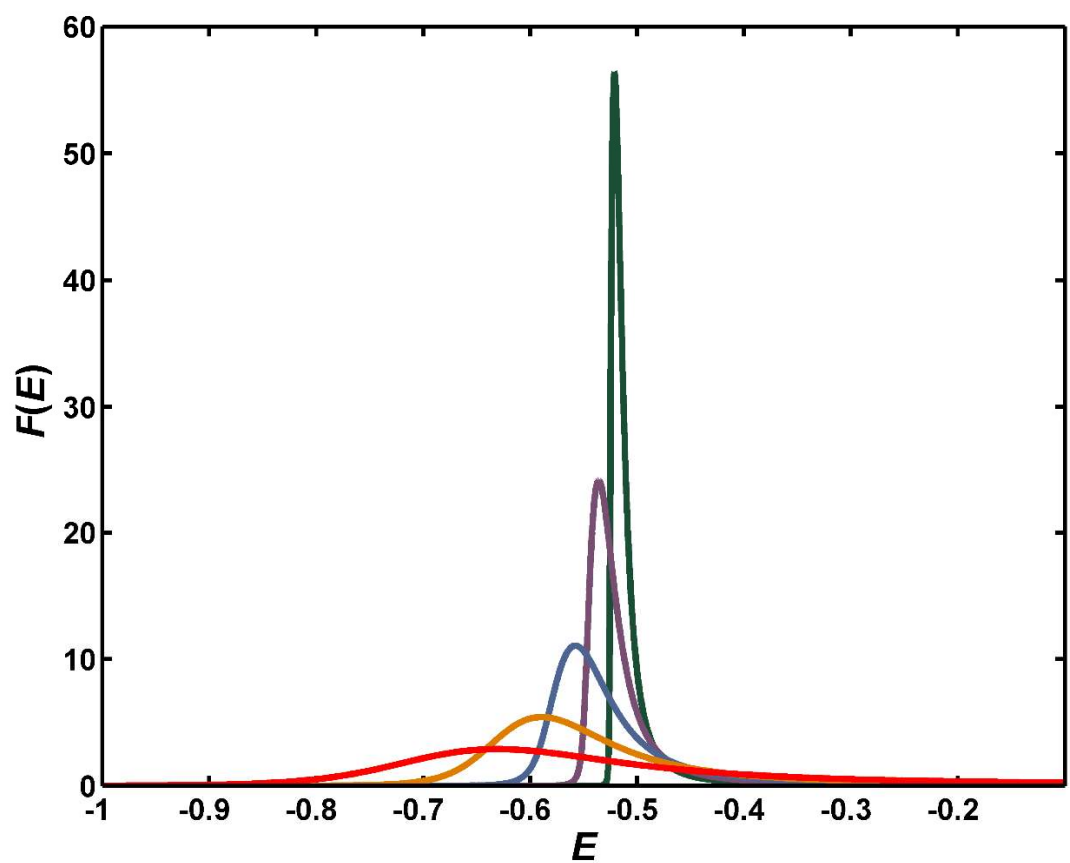

Fig. 1

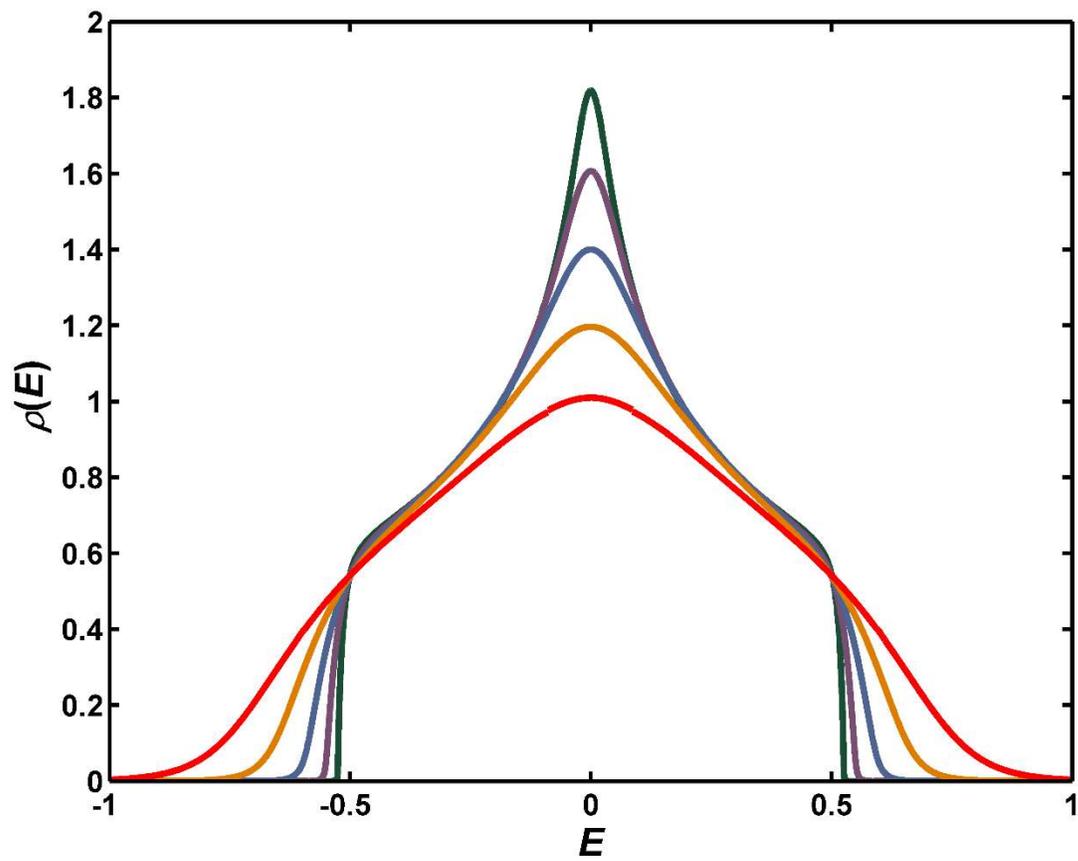

Fig. 2

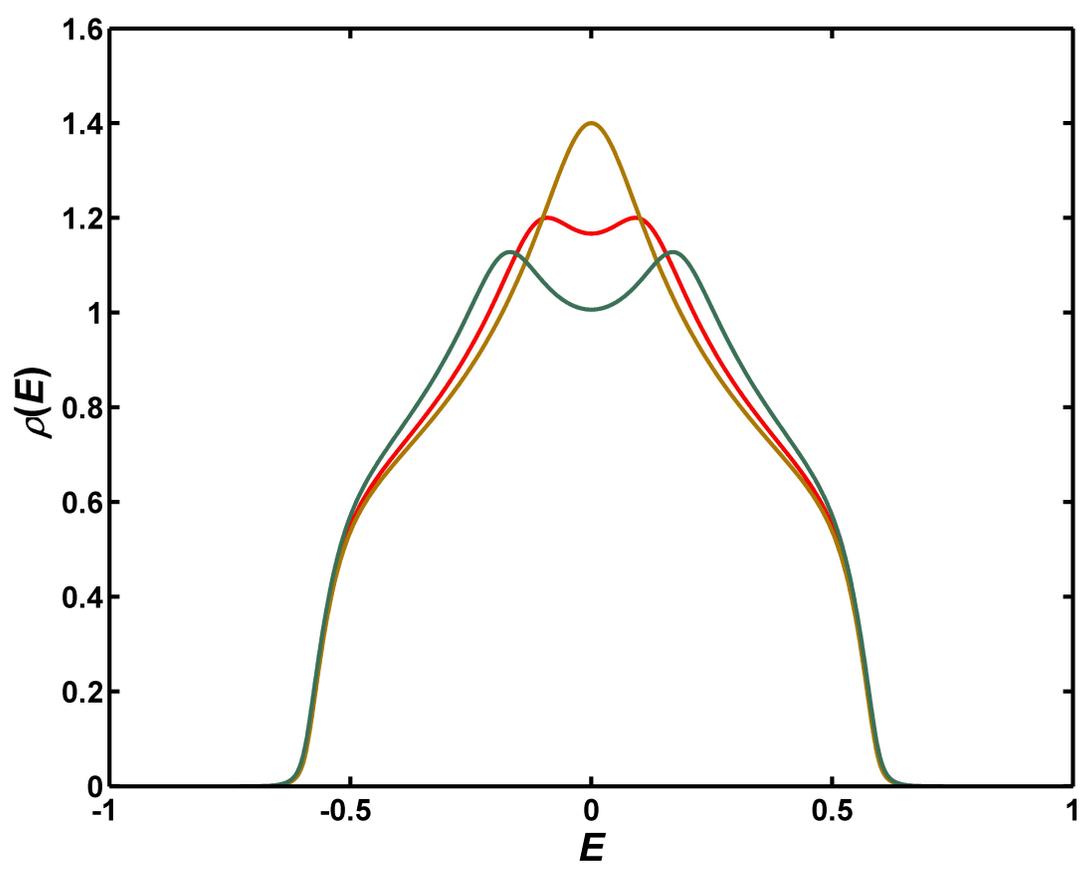

Fig. 3

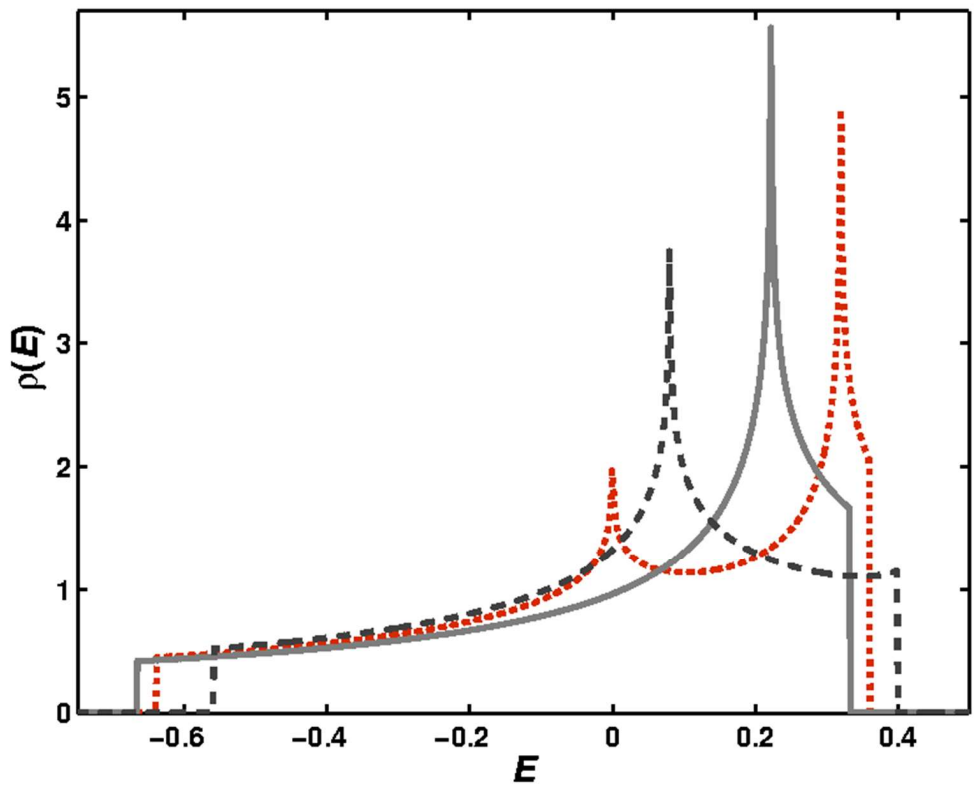

Fig. 4

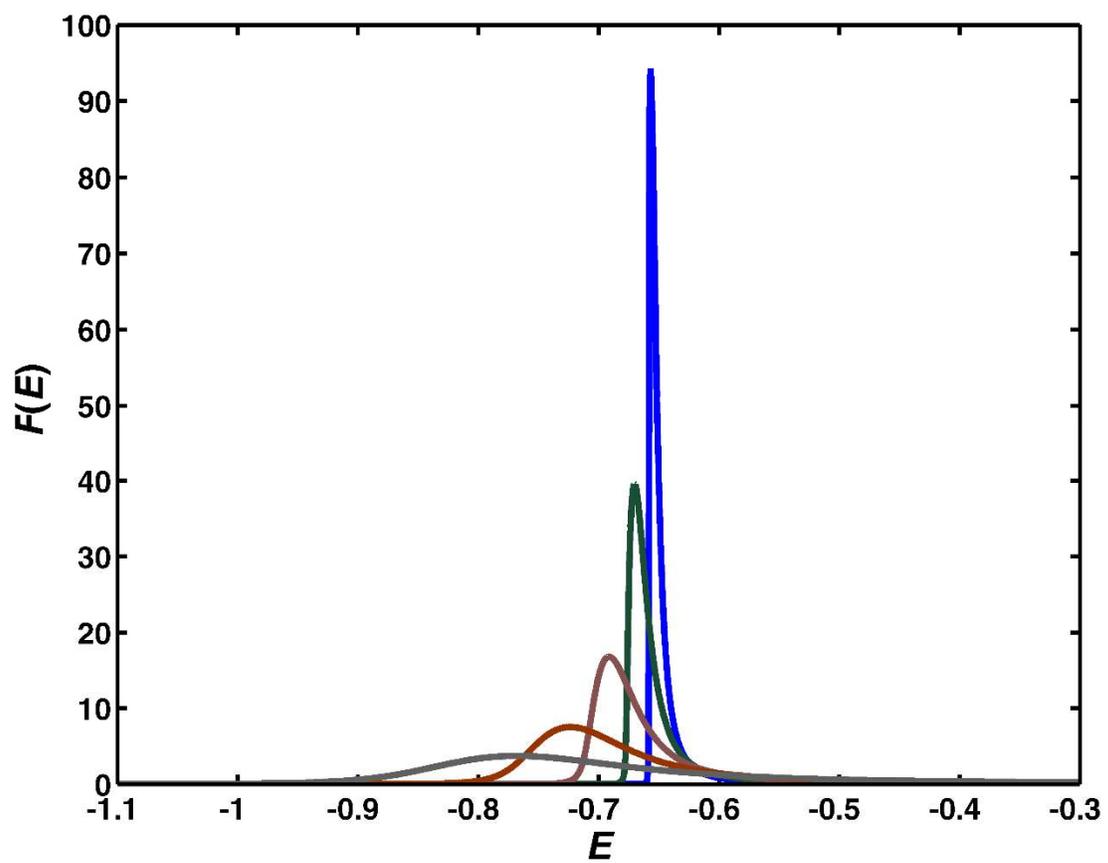

Fig. 5

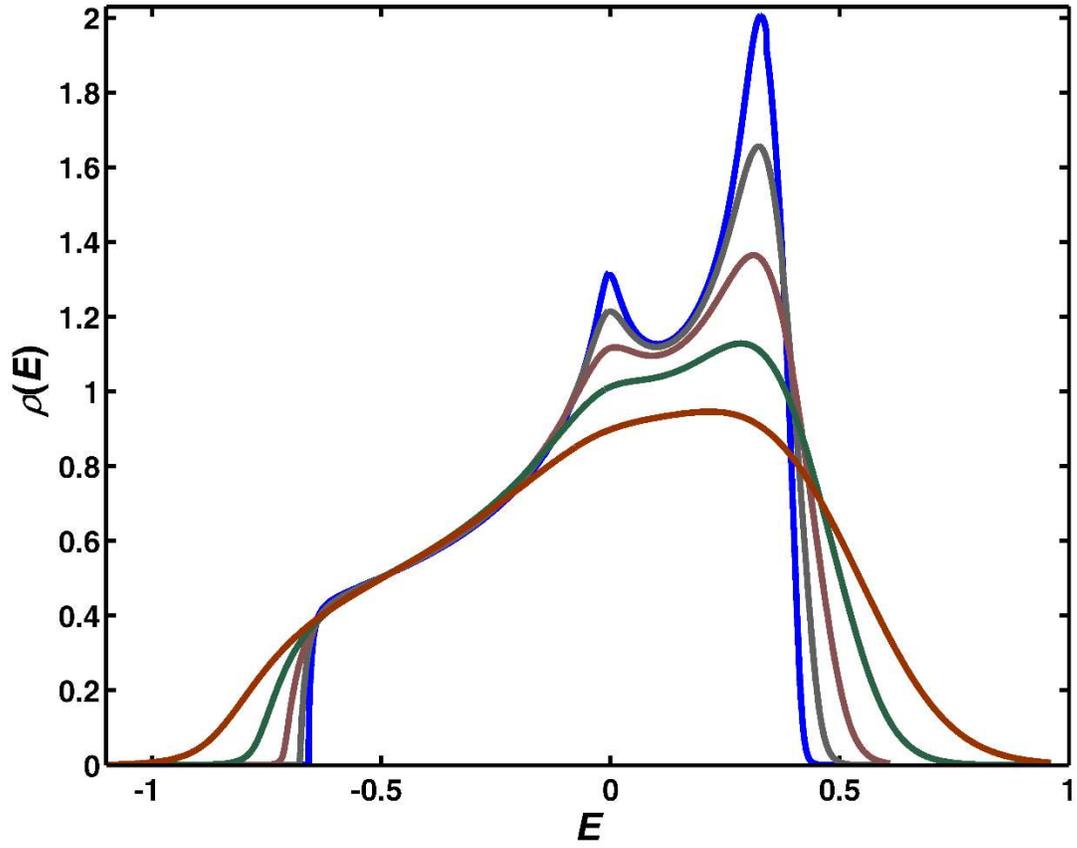

Fig. 6

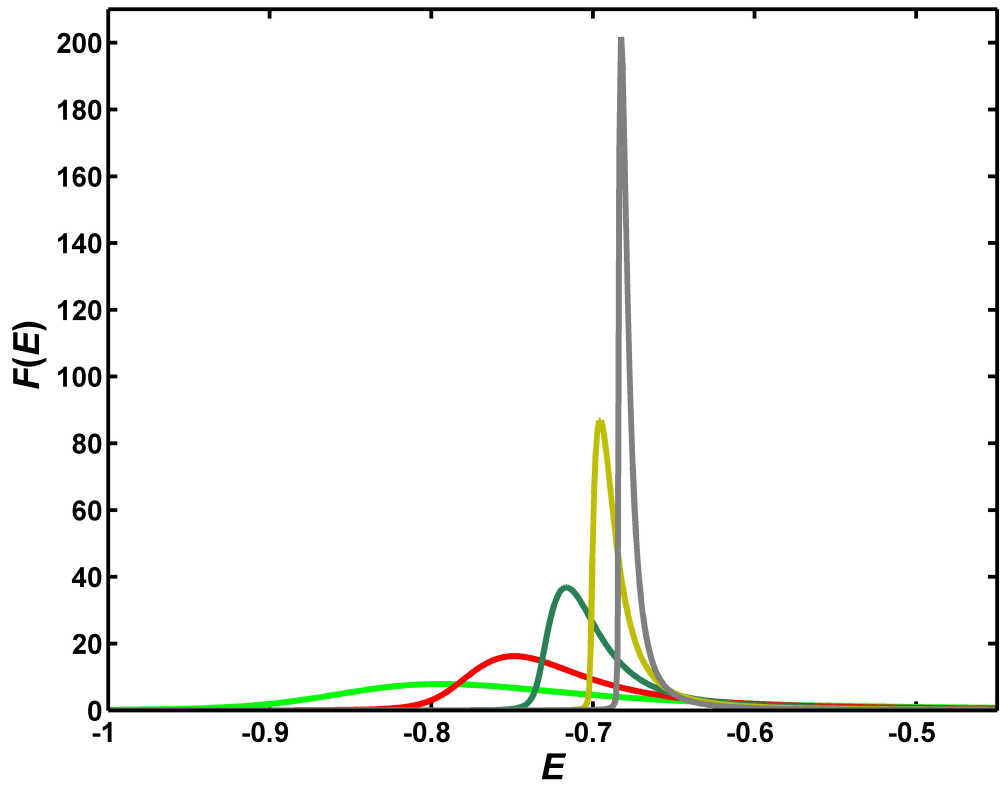

Fig. 7

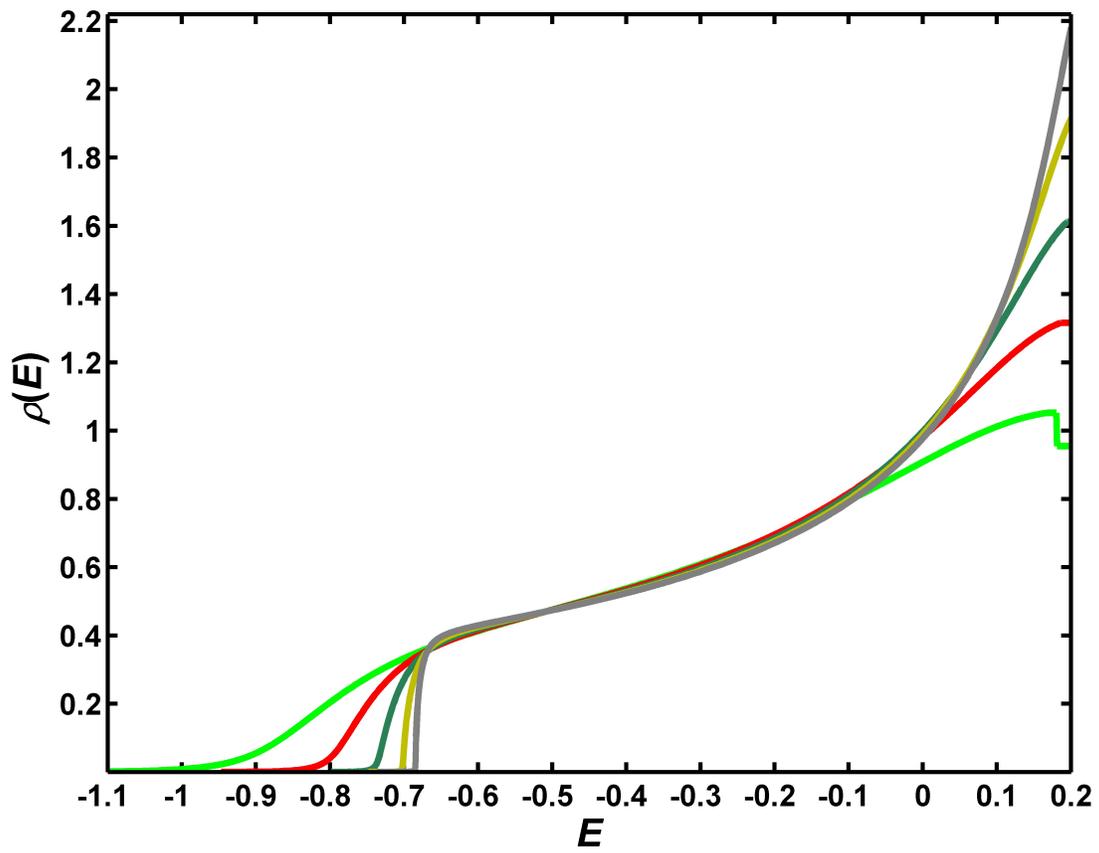

Fig. 8

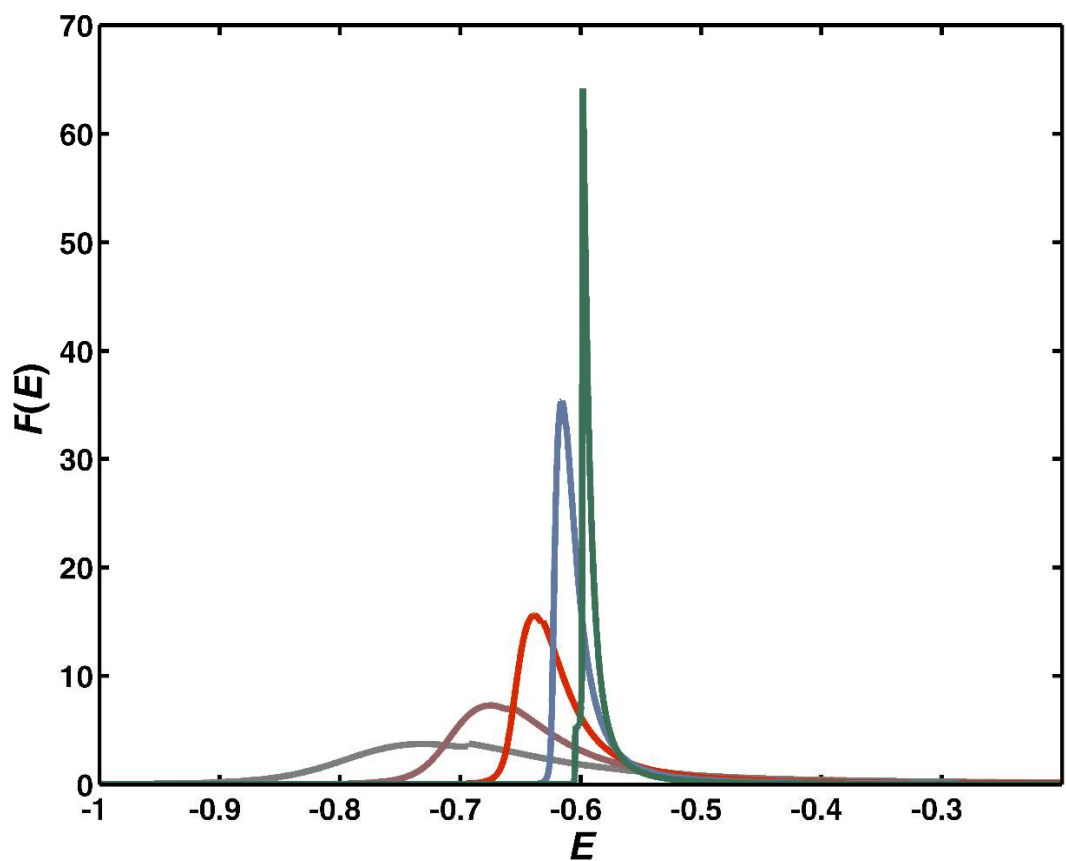

Fig. 9

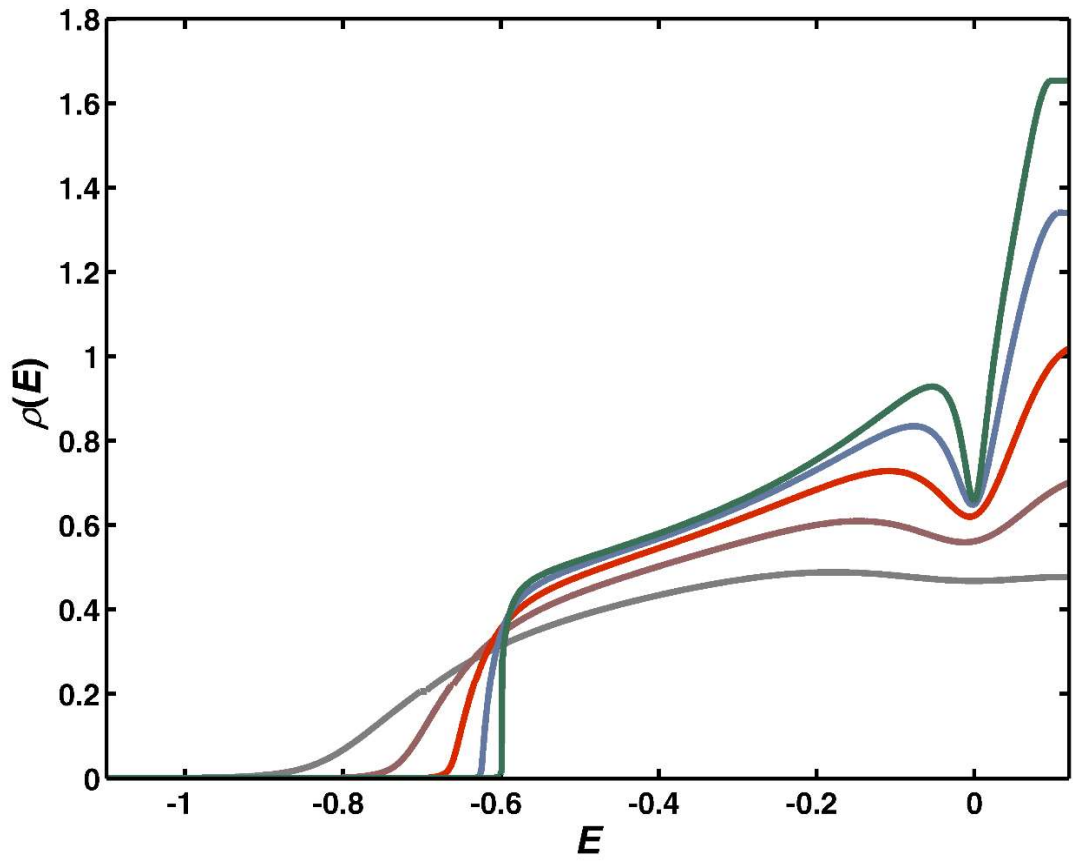

Fig. 10